\documentclass[smallcondensed,natbib]{svjour3}

\usepackage{graphicx}
\graphicspath{{images/}}
\usepackage[caption=false,font=footnotesize]{subfig}

\usepackage{tabularx}
\usepackage{multirow}

\usepackage{tikz}
\usepackage{pgfplots}
\usepgfplotslibrary{statistics}
\pgfplotsset{compat=1.14}
\usetikzlibrary{external}
\usepackage{siunitx}
\usepackage{amsmath}

\usepackage[acronym,shortcuts,nomain]{glossaries}
\newacronym{UAV}{UAV}{unmanned aerial vehicle}
\newacronym{CPS}{CPS}{cyber-physical system}
\newacronym{UGV}{UGV}{unmanned ground vehicle}
\newacronym{UUV}{UUV}{unmanned underwater vehicle}
\newacronym{USV}{USV}{unmanned surface vehicle}
\newacronym{FSM}{FSM}{finite state machine}
\newacronym{ROS}{ROS}{robot operating system}
\newacronym{SysML}{SysML}{systems modeling language}
\newacronym{UML}{UML}{unified modeling language}
\newacronym{SAR}{SAR}{search and rescue}
\newacronym{GPS}{GPS}{global positioning system}
\newacronym{VTOL}{VTOL}{vertical take-off and landing}
\newacronym{ID}{ID}{identifier}
\newacronym{BSD}{BSD}{Berkeley software distribution}
\newacronym{GPL}{GPL}{general public license}
\newacronym{MIT}{MIT}{Massachusetts institute of technology}
\newacronym{UWB}{UWB}{ultra-wideband}
\newacronym{POC}{POC}{proof of concept}
\newacronym{MBSE}{MBSE}{model based system engineering}
\newacronym{BDD}{BDD}{block definition diagram}
\newacronym{IBD}{IBD}{internal block diagram}
\newacronym{FOV}{FOV}{field of view}
\newacronym{VTL}{VTL}{velocity template language}
\newacronym{SCXML}{SCXML}{state chart XML}
\newacronym{BDRML}{BDRML}{behaviour-data relations modeling language}
\newacronym{XML}{XML}{extensible markup language}
\newacronym{JSON}{JSON}{JavaScript object notation}
\newacronym{ZMTP}{ZMTP}{ZeroMQ message transport protocol}
\newacronym{WLAN}{WLAN}{wireless local area network}
\newacronym{IQR}{IQR}{interquartile range}
\newacronym{RC}{RC}{radio control}
\newacronym{API}{API}{application programming interface}
\newacronym{BT}{BT}{behavior tree}
\newacronym{RTF}{RTF}{real time factor}
\newacronym{CPU}{CPU}{central processing unit}

\usepackage[pdfusetitle]{hyperref}

\usepackage{natbib}
\setcitestyle{numbers}

\begin{document}

% *** TITLE ***
%
\title{Engineering Swarms of Cyber-Physical Systems with the CPSwarm Workbench}

% *** AUTHORS ***
%
\author{%
Micha Sende \and
Melanie Schranz \and
Gianluca Prato \and
Etienne Brosse \and
Omar Morando \and
Martina Umlauft
}
\institute{
Micha Sende (corresponding author) \and Melanie Schranz \and Martina Umlauft \at
Lakeside Labs, Klagenfurt, Austria\\
\email{lastname@lakeside-labs.com} \and
Gianluca Prato \at
LINKS Foundation, Turin, Italy\\
\email{gianluca.prato@linksfoundation.com} \and
Etienne Brosse \at
Softeam Research and Development Department, Paris, France\\
{etienne.brosse@softeam.fr} \and
Omar Morando \at
Digisky, Turin, Italy\\
\email{morando@digisky.it}
}

% *** TITLE AREA ***
%
\maketitle

% *** CONTENTS ***
%
\begin{abstract}

% 150 to 250 words: 174 words

% problem statement: 11 words
Engineering swarms of \acp{CPS} is a complex process.
% approach: 50 words
We present the CPSwarm workbench that creates an automated design workflow to ease this process. This formalized workflow guides the user from modeling, to code generation, to deployment, both in simulation and on \ac{CPS} hardware platforms. The workbench combines existing and emerging tools to solve real-world \ac{CPS} swarm problems.

% approach: 76 words
As a proof-of-concept, we use the workbench to design a swarm of \acp{UAV} and \acp{UGV} for a \ac{SAR} use case. We evaluate the resulting swarm behaviors on three levels. First, abstract simulations for rapid prototyping. Second, detailed simulation to test the correctness of the results. Third, deployment on hardware to demonstrate the applicability. We measure the swarm performance in terms of area covered and victims rescued.
% results: 14 words
The results show that the performance of the swarm is proportional to its size.
% conclusion: 23 words
Despite some manual steps, the proposed workbench shows to be well suited to ease the complicated task of deploying a swarm of \acp{CPS}.


\end{abstract}

\glsresetall % reset used abbreviations
\section{Introduction}
\label{sec:intro}

% general issue: challenges in cps software development
Developing software to control an individual \ac{CPS} or even a system of \acp{CPS}  is challenging in multiple ways. First, \ac{CPS} setups typically involve wildly different pieces of hardware running on different platforms~\cite{Rickert2017} with different constraints and requirements. Second, \acp{CPS} are equipped with multiple sensors and actuators to link the physical world with a virtual one~\cite{Reimann2017}. Third, \acp{CPS} are highly connected and thus show a high degree of interaction between individual entities~\cite{Sztipanovits2013} introducing challenges for communication and coordination. Such highly-networked systems become increasingly hard to design which may lead to unpredictable behavior~\cite{Lee2008}.

% background: swarm intelligence --> swarm robotics --> swarms of cps
A possible approach to design and develop such a system of \acp{CPS} is to use the inspiration of swarm intelligence~\cite{Bonabeau1999} where every autonomously acting \ac{CPS} is a member of a large swarm of \acp{CPS}. Each swarm member follows a set of local rules that globally leads to emergent swarm behavior. The concept of swarm intelligence is exemplified by swarm robotics~\cite{Sahin2004} which is a research direction that applies swarm intelligence to groups of robots. In this work we generalize this approach from robotics to \acp{CPS}.

% problem statement: engineering swarms of cps
Engineering swarm intelligence algorithms for \acp{CPS} is very challenging as the desired global swarm behavior is hard to predict from the local rules of the individual \acp{CPS}~\cite{Abbott2006}. In this work we take a bottom-up approach to iteratively design the \ac{CPS} controllers that yield the desired swarm behavior. This is achieved through simulation experiments modeled on different levels of abstraction~\cite{Rand2007}. Hence, a well-defined model of the \acp{CPS} and the desired local behavior is required that can be scaled to the different abstraction levels~\cite{Lee2017}. Such a multi-scale model needs to take into account the engineering of both hardware and software components of the \acp{CPS} as well as the composition of the whole swarm.

% research methodology: engineering process
Therefore, we propose a model-driven engineering process that starts from modeling the \ac{CPS} hardware and behavior and results in behavior code that can be deployed to and executed on the \acp{CPS}. The process is based on a hierarchically organized set of behaviors that together build the complete controller of a single \ac{CPS}. Together with a well-defined library of behavior models, an automated code generation process allows to assemble the behavior of the \acp{CPS} in the form of executable code. Based on the multi-scale modelling approach, it is possible to validate the swarm behavior. This is achieved by an abstract simulation that is used for rapid prototyping of the \ac{CPS} behavior resulting in the desired swarm behavior. For verification purposes, it is then qualitatively compared to the behavior resulting from the code generation process using a detailed simulation. These results are used to verify that the model and the generated code yield the behavior previously observed with an abstract simulation. The resulting behavior is demonstrated on hardware in the form of a proof-of-concept.

% motivate sar use case
To prove the practicability of this engineering process, we base our experiments on a \ac{SAR} scenario. This use case envisions a heterogeneous swarm of \acp{UAV} and \acp{UGV} that is deployed in a highly dynamic disaster environment to support first responders in real-time. Its mission is to collectively search for victims (\acp{UAV}), and rescue them (\acp{UGV}). Such a mission is well suited for testing swarms of \acp{CPS} because it can benefit from several swarm properties. The scalability of a swarm allows to add more \acp{CPS} during the mission in case the area to be covered is larger than expected. The robustness of a swarm takes into account challenging environments where individual \acp{CPS} can fail. Moreover, in contrast to fully centralized control, such a swarm can still operate even if connectivity is sparse. The swarm operates autonomously as a self-organized mixed team where the decisions to fulfil the mission are made as appropriate and demanded from the dynamic environment.

% relation to cpswarm
This work is part of the CPSwarm project~\cite{Bagnato2017}. It aims at developing a tightly integrated toolchain that covers the whole design process of \ac{CPS} swarms, starting from modeling and code generation to deployment, running, and monitoring the swarm in simulations and real-world scenarios. This is enabled by the CPSwarm workbench~\cite{Jdeed2019}, a toolchain that has been established to serve as a glue among the individual components in a \ac{CPS} design process~\cite{Gunes2014}. In this paper we highlight parts of this toolchain including modeling, code generation, and execution of the swarm applied to the \ac{SAR} use case. For further details on the workflow, the reader is referred to~\cite{Bagnato2017}.

% main contributions
Our main contributions are:
\begin{itemize}
\item We formalize the engineering process of designing a swarm of \acp{CPS}.
\item We provide a set of tools that are tightly linked to allow \ac{CPS} software development.
\item We provide a library of functions and behaviors to speed up the process of \ac{CPS} swarm software development.
\item We demonstrate the complexity of the engineering process that arises from the deployment onto a swarm of \ac{CPS} hardware platforms. 
\end{itemize}

% outline
After reviewing the state of the art in Section~\ref{sec:related}, we provide a description of the proposed engineering process in Section~\ref{sec:method}. We then exemplify this methodology on the \ac{SAR} use case in Section~\ref{sec:use_case}. To complete the process we perform an evaluation of the resulting behavior in Section~\ref{sec:evaluation} that is used to validate the process. We conclude the paper in Section~\ref{sec:conclusion}.

\section{Related Work}
\label{sec:related}

The CPSwarm toolchain is dedicated to the design process of \acp{CPS} with a focus on a swarm applications. A very similar approach taken by the OpenADx working group~\footnote{\url{https://wiki.eclipse.org/OpenADx}}, which aims at developing a toolchain that targets autonomous driving use cases. It is developed as open source project within the scope of the Eclipse Foundation and comprises similar design steps, namely architecture definition, simulation, integration, test drives, and connectivity-based validation. The initiative spans across multiple industries and started in 2019, thus the toolchain is conceptually on a very early stage. To the best of our knowledge, this is the only other toolchain that integrates the complete design cycle of \ac{CPS} development.

\subsection{Hardware Modeling}
Modeling \ac{CPS} hardware allows to focus on the application by reducing complexity. In the scope of this paper, we work with the concepts of \ac{MBSE}. Models are created with use-case-specific requirements including functional, performance, and interface requirements~\cite{Friedenthal2014}. This allows to design a system by concentrating knowledge and guiding the implementation process.

The INTO-CPS project focuses on \ac{MBSE} for \ac{CPS}~\cite{Larsen2016}. It employs several tools to cover the entire modeling life cycle. Besides sensors and actuators, its \ac{CPS} model consists of the behavior that is modeled as additional ``hardware'' component. This allows sensors and actuators to directly interact with the core behavior. The OpenMETA project follows a similar approach~\cite{Sztipanovits2014}. It provides a model and component-based design toolchain for complex \acp{CPS} that supports model, tool, and design process integration. The CERBERO project~\cite{Masin2017} has a cross-layer model-based approach to describe, optimize, and analyze energy efficient and secure \acp{CPS} focusing on adaptivity. It uses data models that allow reconfiguration at runtime. A comprehensive overview of modeling technologies for \acp{CPS} is given in~\cite{Quadri2015}. Hardware models may also be combined with behavior models as proposed in~\cite{Fitzgerald2014}.

Our work applies \ac{MBSE} approach proposed in the INTO-CPS project. We extend the existing work by allowing to model not only the hardware composition of a \ac{CPS}, i.e., the sensors, actuators, and controller, but also the composition of a swarm of \acp{CPS}. The swarm composition describes the type and cardinality of \acp{CPS} included in the swarm.

\subsection{Behavior Modeling}
The behavior of swarm agents is typically modeled by textual and mathematical formulations~\cite{Hamann2008,Hassanien2015,Yang2018},
pseudo code~\cite{Parpinelli2011}, or diagrams~\cite{Hamann2008}. Especially the latter is experiencing a high popularity since diagrams can be represented both visually to be read by humans and formally for automatic processing. A prominent example are \acp{FSM} which have been used to model the behavior of individual robots~\cite{Brooks1986,deAraujo2014}, networked systems~\cite{Spears2007}, multi-robot systems~\cite{Rabbath2013}, and swarm robotic systems~\cite{Soysal2005}. Other approaches go further by defining a new language for specific purposes. For example, Modelica is a language for modeling and simulation of physical systems~\cite{Fritzson2006}. It has been applied to \ac{CPS} simulation in~\cite{Henriksson2011}. The \ac{BDRML} introduced in~\cite{Pitonakova2017} focuses on modeling the information exchange between robots in a swarm.

More recently, \acp{BT} have been adopted from the game industry and applied to swarm robotics behavior modeling~\cite{Colledanchise2017,Kuckling2018}. They pick up the idea of the subsumption architecture~\cite{Brooks1986} which proposes several abstraction layers to simplify the design process. This approach is similar to our proposal of hierarchical \acp{FSM} with the main difference that behaviors are assembled from tasks rather than states.

A less formal approach are design patterns that have been successfully instrumented for taking design decisions in recurring problems. Design patterns are developed over time by a community of experts. They are usually formulated in prose to avoid specific vocabulary and to keep the formulations understandable for all stakeholders that are involved in the design process. Recent research takes into account semi-formal and formal approaches in order to automate the structuring, retrieval, and selection processes for patterns using specific libraries~\cite{Cornils2000,Eden1997}. Similarly, other types of formal behavior models can be placed in libraries~\cite{McLurkin2004}. This enables their reuse and the creation of new behaviors by combining existing models~\cite{Do2003,FernandezMarquez2013}.

However, a common representation for swarms of \acp{CPS} is still missing. Therefore, we propose a modeling approach that is flexible enough to feature different complex scenarios using different types of hardware. This approach builds on the well established concepts of \ac{UML}, \ac{SysML}, and \acp{FSM} to increase the acceptance.

\subsection{Code Generation}
Code generation can be tackled on multiple levels. Most commonly, a template based code generation~\cite{Kelly2008} is used which enables generation of executable code from behavior models based on pre-defined templates. In the domain of \acp{CPS}, there are approaches for automated controller synthesis~\cite{Roy2011,Martin2012} which generate complete \ac{CPS} controllers from abstract \ac{CPS} models. Furthermore, there are approaches that provide comprehensive tool chains to integrate the modeling and \ac{CPS} controller synthesis~\cite{Brosse2012,Larsen2016}.

The CPSwarm workbench provides a step forward in terms of simplification and correct code generation of \ac{CPS} behavior, enabling an almost fully automated approach that avoids the introduction of errors in manual implementation of \ac{CPS} behavior software. The CPSwarm code generator does not aim to substitute developer work but to support the development by simplifying the process of integrating \ac{CPS} swarm modeling with the deployment of \ac{CPS} controllers. In particular, this objective is achieved by integrating the template based code generation techniques with an abstraction layer into the CPSwarm workbench.

\section{Engineering CPS Swarms}
\label{sec:method}

\begin{figure*}
\centering
\includegraphics[width=\linewidth]{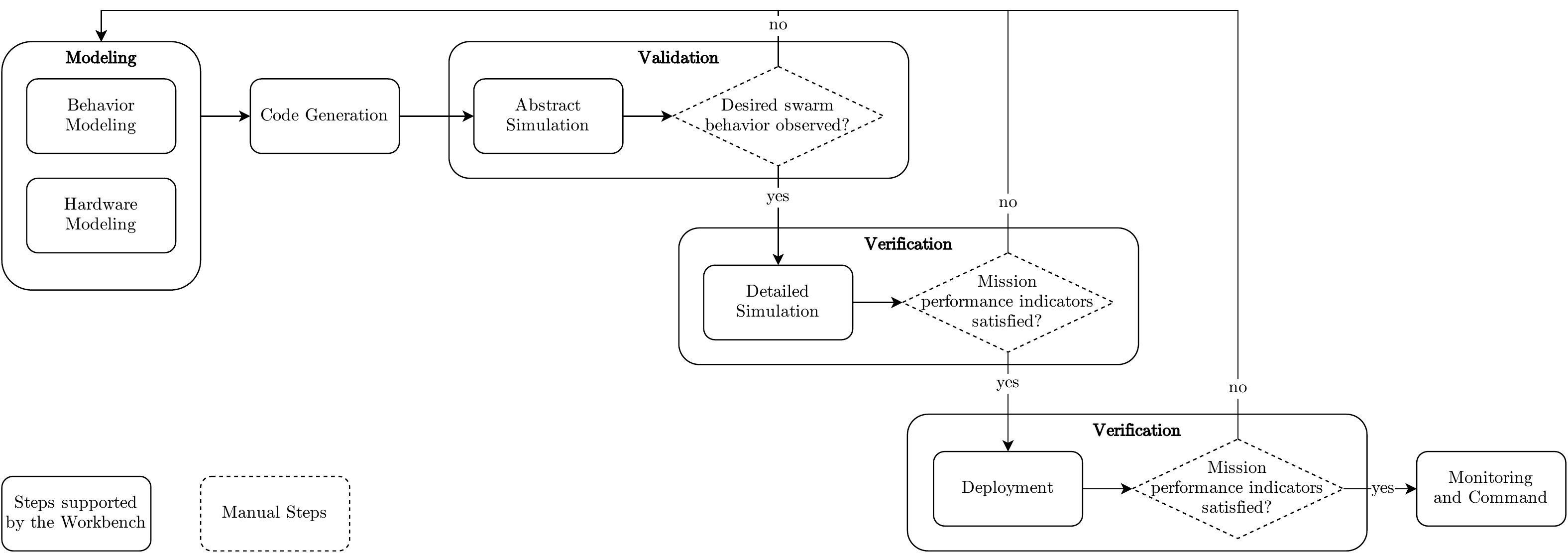}
\caption{The proposed workflow for engineering swarms of \acp{CPS}.}
\label{fig:engineering_process}
\end{figure*}

\begin{figure*}
\centering
\includegraphics[width=\linewidth]{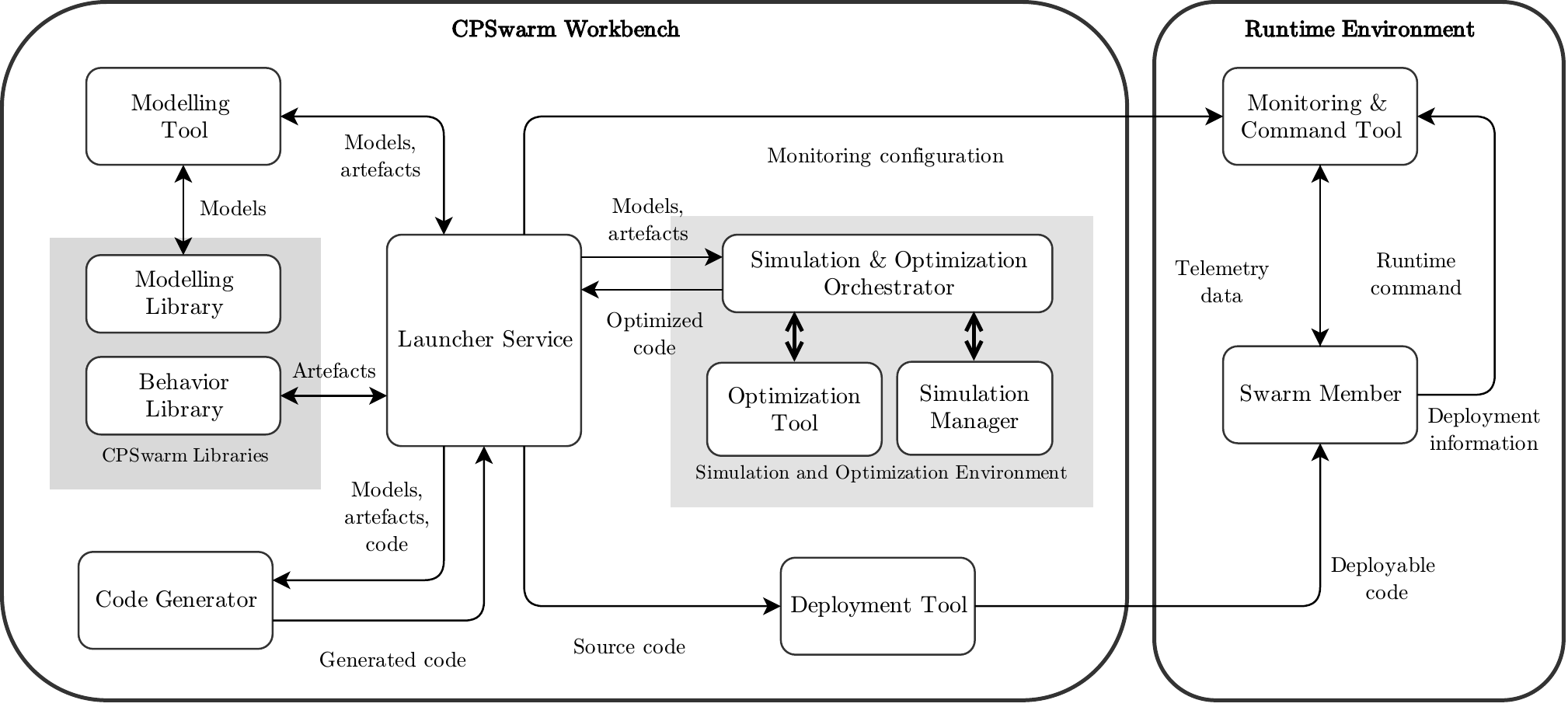}
\caption{The CPSwarm workbench architecture showing the different tools, their interconnections, and the type of information flowing between the tools.}
\label{fig:workbench_architecture}
\end{figure*}

Engineering a swarm of \acp{CPS} comes with a number of decisions. This includes (i) deployment specifics such as number, type, and placement of \acp{CPS}, (ii) hardware of individual \acp{CPS}, e.g., communication interface, on-board sensors, or battery dimensioning, and (iii) required behavior of individual \acp{CPS} and the \ac{CPS} swarm. In this section we describe our engineering process including \ac{CPS} modeling and automatic \ac{CPS} code generation for simulation and deployment. This process is visualized by the activity diagram in Figure~\ref{fig:engineering_process}.
It is supported by the CPSwarm workbench whose architecture is visualized in Figure~\ref{fig:workbench_architecture}.
The architecture diagram shows the actual interconnections between the different tools that are used at each step. It can be seen that all data is passed through and handled by the central launcher. All tools of the workbench are released as open source software packages in the CPSwarm GitHub repository~\footnote{\url{https://github.com/cpswarm/workbench}}. In the following we discuss modeling and code generation. The validation and verification process is detailed in Section~\ref{sec:evaluation}.

\subsection{Hardware Modeling}
One main aspect of modeling a \ac{CPS} consists of specifying its architecture in terms of hardware components and their interactions. We build the hardware models using \ac{SysML}~\cite{Arnould2017}. It extends \ac{UML} by putting a focus on modeling systems with block diagrams. To enable modeling of \ac{CPS} swarms, we built a \ac{CPS} swarm profile~\cite{Schranz2019} extending the \ac{BDD} and \ac{IBD}. It includes three types of diagrams:
\begin{itemize}
\item The swarm composition diagram extends the \ac{BDD} that allows to model the swarm as a whole. It specifies the types and number of \acp{CPS} used in the swarm.
\item The hardware composition diagram extends the \ac{BDD} to model the individual hardware components used by the \acp{CPS}. It specifies which types of hardware components exist and defines their parameters such as inputs and outputs.
\item The swarm member internal diagram extends the \ac{IBD} to model the structure of a single \ac{CPS}. It specifies how the behavior algorithms can interact with the world by defining sensors, actuators, and communication interfaces. Where sensors are inputs to collect information from the environment and actuators are outputs to interact with the environment, a communication interface can be both input and output to the swarm algorithms and enables coordination between \acp{CPS}.
\end{itemize}

\subsection{Behavior Modeling}
The behavior models define the software components of each \ac{CPS}. They are modeled to describe how the \ac{CPS} behaves by interacting with the environment and the other \acp{CPS} in the swarm. The behavior is defined in a way that the \ac{CPS} swarm executes and completes the mission that is intended by the designer of the swarm system. Designing the behavior of individual agents in a swarm is a difficult task because the emergent swarm behavior is not easily predictable~\cite{Abbott2006}. Modeling the behavior on an abstract level facilitates the process by allowing to execute it on different levels of realism. This bottom-up approach allows to iteratively refine the individual behaviors until a global swarm behavior is reached. The formal behavior models allow to speed up the design process by automatically generating the code to be executed on the \acp{CPS}, either in simulation or in real-world experiments.

\subsubsection{Behavior State Machines}
% state machines
A mission for a swarm of \acp{CPS} typically requires many different behaviors to be executed by the \acp{CPS} to complete the different tasks of the mission. These individual behaviors can be simpler to describe and implement, and are regarded to be atomic during modeling. Combining these simple behaviors into more complex behaviors allows to achieve complex missions. A commonly used approach for this are \acp{FSM}, where the states correspond to simple behaviors and the \ac{FSM} describes a complex behavior. Each \ac{CPS} can execute an \ac{FSM} while always being in a defined state which can vary between \acp{CPS}. Hence, complex swarm configurations can emerge where \acp{CPS} take on different roles based on their interactions.

% events
The transitions between behavior states are triggered by events that can either originate locally, e.g., from sensor readings or behavior rules, or remotely, e.g., from communication between \acp{CPS} or commands from a control station. Exchanging events between \acp{CPS} enables the coordination of the swarm behaviors. The way events are processed locally is done autonomously by each \ac{CPS} and defined in the behavior \acp{FSM}. This allows \acp{CPS} to influence each other's behavior changes by exchanging events. Events are formalized by a unique \ac{ID}, a timestamp, and a unique \ac{ID} of the sender. Furthermore, events can have data associated with them that is passed among behaviors.

% behavior types
The behavior \ac{FSM} model is based on the \ac{UML} behavior \acp{FSM}~\cite{Cook2017}. Specifically, the simple behaviors are modeled by simple states and the complex behaviors are modeled using composite or submachine states. Composite states allow a state to be modeled by another \ac{FSM} whereas submachine states allow to encapsulate generic \acp{FSM} that can be reused within more than one state. In the context of \ac{CPS} swarm behaviors, we propose four different behavior types (the color codes refer to the ones used in Figure~\ref{fig:behavior_library}):
\begin{itemize}
\item Complex behaviors (red) are defined by an \ac{FSM} of simple behaviors, e.g., \ac{SAR}.
\item Swarm behaviors (green) are simple behaviors that execute a specific swarm algorithm exhibiting an emergent swarm behavior, e.g. aggregation or exploration.
\item Swarm functions (blue) are simple behaviors that execute a single function including the interaction between \acp{CPS}, e.g. task allocation or exchanging position information.
\item Hardware functions (yellow) are simple behaviors that execute a single function including hardware interaction, e.g., moving to a given location or controlling an actuator.
\end{itemize}
The behaviors are formalized by a unique name, a short description of the behavior, the behavior type, and the inputs and outputs of the behavior.

% state machine hierarchy
In related literature there typically does not exists such a fine grained distinction between different behavior types. For example, swarm behaviors and swarm functions are termed collective behavior in~\cite{Brambilla2013} and basic swarm behavior in~\cite{Schranz2020}. In~\cite{Francesca2014}, constituent behavior is used as a term for any kind of simple behavior including hardware functions. We deliberately introduce the distinction between the behavior types in order to foster  clean and structured design of the behavior \acp{FSM}. This is enabled through hierarchically nested states as defined in the \ac{UML} standard. This allows to abstract behavior details at higher levels. For example, an aggregation behavior can thus be implemented agnostic to the way position information is exchanged between \acp{CPS}. Events can either trigger a state change of the sub \ac{FSM} or force the higher-level behavior itself to terminate and thereby also terminate the currently running sub behavior.

% TODO: unify capitalization of library / behavior names

\subsubsection{Behavior Libraries}
\begin{figure}
\centering
\includegraphics[width=\linewidth]{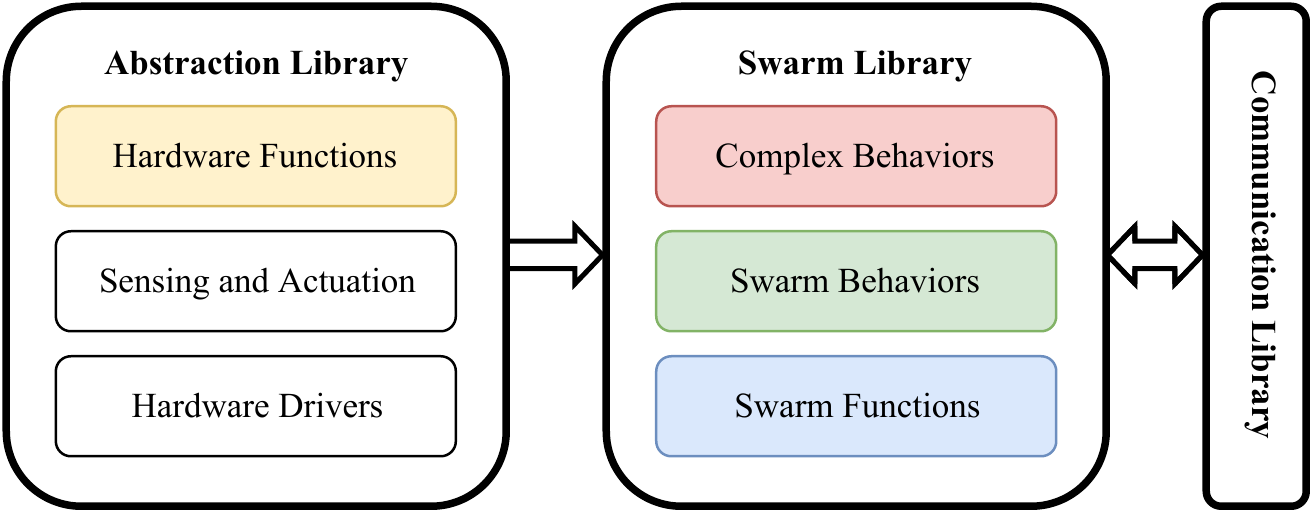}
\caption{The behavior library structure.}
\label{fig:behavior_library}
\end{figure}

The behaviors are placed in libraries to allow frequently recurring behaviors and functionalities to be defined only once. We propose the library structure shown in Figure~\ref{fig:behavior_library} where the colors represent the different behavior types previously introduced.
These libraries are structured according to the level of hardware abstraction to enable a separation of concerns. They contain software artifacts that are modeled as states of the \acp{FSM}. The Swarm Library works independently of the underlying hardware. It provides the complex behaviors modeled as \acp{FSM} together with the swarm behaviors and swarm functions. It makes use of a Communication Library that provides an interface for communication among \acp{CPS}. The Abstraction Library abstracts away the hardware specifics. It provides functions that are related to the hardware and functionalities to access sensors and actuators based on hardware specific drivers.

\paragraph{The Swarm Library} contains the swarm behaviors executed by the individual \acp{CPS} leading to the global behavior of the swarm. They are platform independent and thus can be reused among different types of \acp{CPS}. The Swarm Library is structured into three sub libraries in accordance with the previously introduced swarm behaviors. First, the Complex Behaviors library contains the \acp{FSM} that model the complex, high-level mission behaviors such as \ac{SAR}. They are defined as \ac{UML} composite or submachine states. Second, the Swarm Behaviors library contains individual swarm algorithms that exhibit an emergent behavior. Typically, such swarm algorithms are handcrafted based on biological inspiration or generated automatically, e.g., using evolutionary optimization. Examples are flocking, phototaxis, or collective transport. They are defined as \ac{UML} simple states to be used in the complex behavior \acp{FSM}. Third, the Swarm Functions library contains simple swarm related tasks. These are tasks that do not lead to an emergent behavior but rather are used to enable the functioning of the swarm behaviors. Examples are exchange of position information, task allocation, or computing the average velocity of the swarm. They are defined as \ac{UML} simple states to be used in the complex behavior \acp{FSM}.

\paragraph{The Communication Library} provides communication services to the swarm that are built on an arbitrary network interface. These services include transmission of telemetry from \acp{CPS} to the command and control station, exchange of events between \acp{CPS} and between \acp{CPS} and the command and control station, and remote access to parameters of the \acp{CPS} in the swarm. The Communication Library developed as part of this work is released on GitHub\footnote{\url{https://github.com/cpswarm/swarmio}}.

\paragraph{The Abstraction Library} allows to access the hardware of the \ac{CPS}. It raises the level of abstraction from a platform-dependent point of view to an application-oriented perspective to facilitate the development of high-level routines. This allows to concentrate on describing how the \acp{CPS} should behave in order to complete a high-level task or reach an application-specific goal. This is achieved by providing a set of \ac{CPS}-specific libraries in order to access platform-specific information of a \ac{CPS} in a standard and coherent way.

To achieve this, the Abstraction Library is organized as a composition of three layers where each layer adds a level of hardware abstraction. First, the bottom most layer of Hardware Drivers gathers the software libraries that are responsible for enabling the other layers to access the hardware functionalities. This layer constitutes the foundation of the Abstraction Library and includes all the drivers for sensors and actuators that are mounted on the \acp{CPS}. Second, the Sensing and Actuation layer is responsible for providing sensor information and for controlling the \acp{CPS} using their actuators. While the Hardware Drivers layer has a direct connection with the hardware, this layer purely consists of software that contributes to supply a first degree of abstraction by realizing complex functionalities required by the overlying layer. Finally, the topmost layer of Hardware Functions exhibits a set of high-level functions corresponding to complex routines that a \ac{CPS} can execute involving a set of sensors and actuators. Each function interacts with the lower layers for sending actuator commands and requesting sensor information. These functions constitute the base building blocks to define the states of the \acp{FSM} as \ac{UML} simple states.

\subsection{Code Generation}
The code generator is responsible for transforming the \ac{FSM} models into code that can be deployed and executed on the \acp{CPS}. The generation is template-based which is well suited for such schematic and repeatable structures. Template-based generation defines a simple set of target-templates to be filled with data extracted from the algorithm specification.

These templates are text files, composed of a static part that appears one-to-one in the output and a dynamic part replaced by input data, written in a template meta-code. This code contains all the directives that are processed together with the input data by a template engine to produce the final source code. Different templates allow to target different runtime platforms. A set of templates to generate executable code is realized using the \ac{VTL}.

The data that is processed by the code generator to produce the output are the \ac{FSM} models of the complex behaviors. When the modeling phase is completed, the resulting \ac{FSM} is translated to a \ac{SCXML} file. The resulting file contains all the information needed by the code generator to complete its task. In particular, it contains the information related to the type of the functionality that is used to select the correct template during the generation process. Furthermore, it contains the definition of the \ac{API} of the corresponding functionality that is executed in a specific state.

Using the \ac{SCXML} file and the set of templates, the code generator can produce Python code, implementing the designed \ac{FSM} using the SMACH library. SMACH is a Python-based project, that allows easy implementation and execution of \ac{FSM}-based algorithms. The code generator developed as part of this work is released on GitHub\footnote{\url{https://github.com/cpswarm/code-generator}}.

\section{Use Case: Search and Rescue}
\label{sec:use_case}

In this section, we demonstrate the previously proposed formal approaches using a heterogeneous swarm of \acp{CPS}. We describe how we model the hardware and software components that are used for the experimental evaluation. The focus is to demonstrate the feasibility of deploying the software onto \acp{CPS} from the models. We use the modeling tool Modelio to create these diagrams and then export the data to be further processed by the code generator using the CPSwarm Modeler module\footnote{\url{https://github.com/cpswarm/modelio-cpswarm-modeler}}.

For this demonstration, we choose the use case of \ac{SAR}. In such missions, \acp{CPS} can be used for multiple tasks including creation of a situational overview, logistic support, acting as repeater or surrogate for other \acp{CPS}, and removal of rubble~\cite{Murphy2008}. As our focus is on the demonstration of the workbench including deployment on hardware platforms, we simplify the \ac{SAR} scenario in such a way that it can be accomplished with prototype platforms that do not require specialized hardware. Hence, the tasks to be completed by the \acp{CPS} is finding human casualties or persons trapped in the disaster site and delivering first aid. We implement the heterogeneous swarm using \acp{UAV} and \acp{UGV}.
The mission of the \acp{CPS} swarm can be described as follows: the \acp{UAV} cover a defined environment searching for victims while the \acp{UGV} stay idle. Once a \ac{UAV} discovers a victim, it switches from coverage to tracking. It continuously tracks the victim position which it communicates to the \acp{UGV}. To select the most appropriate \ac{UGV} to reach the victim, the \ac{UAV} starts an arbitration process. The \ac{UGV} is selected based on a cost value submitted by the \ac{UGV}, e.g., the distance to the victim. The selected \ac{UGV} navigates to the victim using the position received from the \ac{UAV}. After having reached the victim, the victim is assumed to be rescued, the \ac{UGV} returns to its starting position, and the \ac{UAV} restarts the coverage process to find a new victim.

In the following we focus exemplarily on modeling the \acp{UAV}, i.e., the hardware models and the behavior implementation which is based on \ac{ROS}~\cite{Quigley2009}. We omit the simpler \ac{UGV} models for the sake of saving space. We slightly abstract the \ac{SAR} use case by referring to victims as targets.

\subsection{Hardware Models}
\begin{figure*}
\centering
\includegraphics[width=\linewidth]{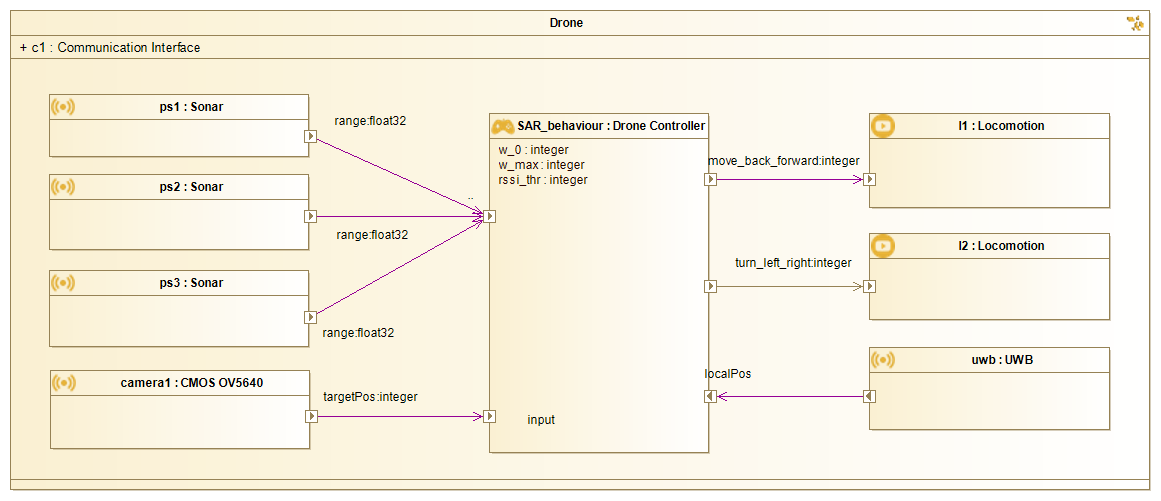}
\caption{The hardware model of the \ac{UAV} prototype.}
\label{fig:hw_model}
\end{figure*}

The hardware platforms used in this work are prototype platforms. The \ac{UAV} platform is custom made by the authors. The \ac{UGV} platform is based on an off-the-shelf \ac{RC} truggy. The extended \ac{SysML} \ac{IBD} model corresponding to the \ac{UAV} prototype platform is shown in Figure~\ref{fig:hw_model}.
The model specifies inputs (sonar range finders, camera, and \ac{UWB} localization), outputs for controlling the locomotion, and a communication interface for communication between the \acp{CPS} and the command and control station. Details about this modeling process can be found in~\cite{Schranz2018}.

\subsection{Behavior Implementation}
To model the behavior of the \acp{UAV}, we use a two-leveled \ac{FSM} hierarchy $H=\{L_1,L_2\}$. It models the swarm behaviors using a set of swarm and hardware functions from different libraries. The implementation of the behavior libraries is based on \ac{ROS} which is an open-source framework to ease the development of robot software. It is currently considered a de facto standard in the robotic community because of its great flexibility and support for a wide variety of hardware platforms.

\begin{figure}
\centering
\subfloat[The \ac{UAV} \ac{SAR} mission processes.]{%
\label{fig:fsm_l1}%
\includegraphics[width=0.41\linewidth]{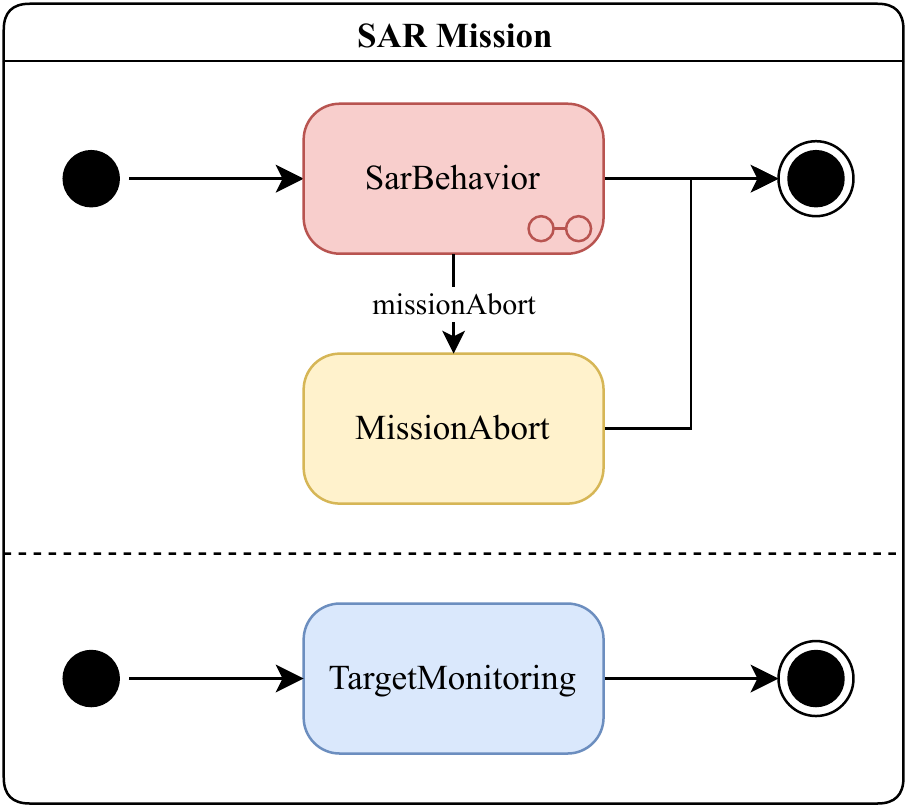}}%
\hfill
\subfloat[The complex \ac{SAR} behavior of the \acp{UAV}.]{%
\label{fig:fsm_l2}%
\includegraphics[width=0.5\linewidth]{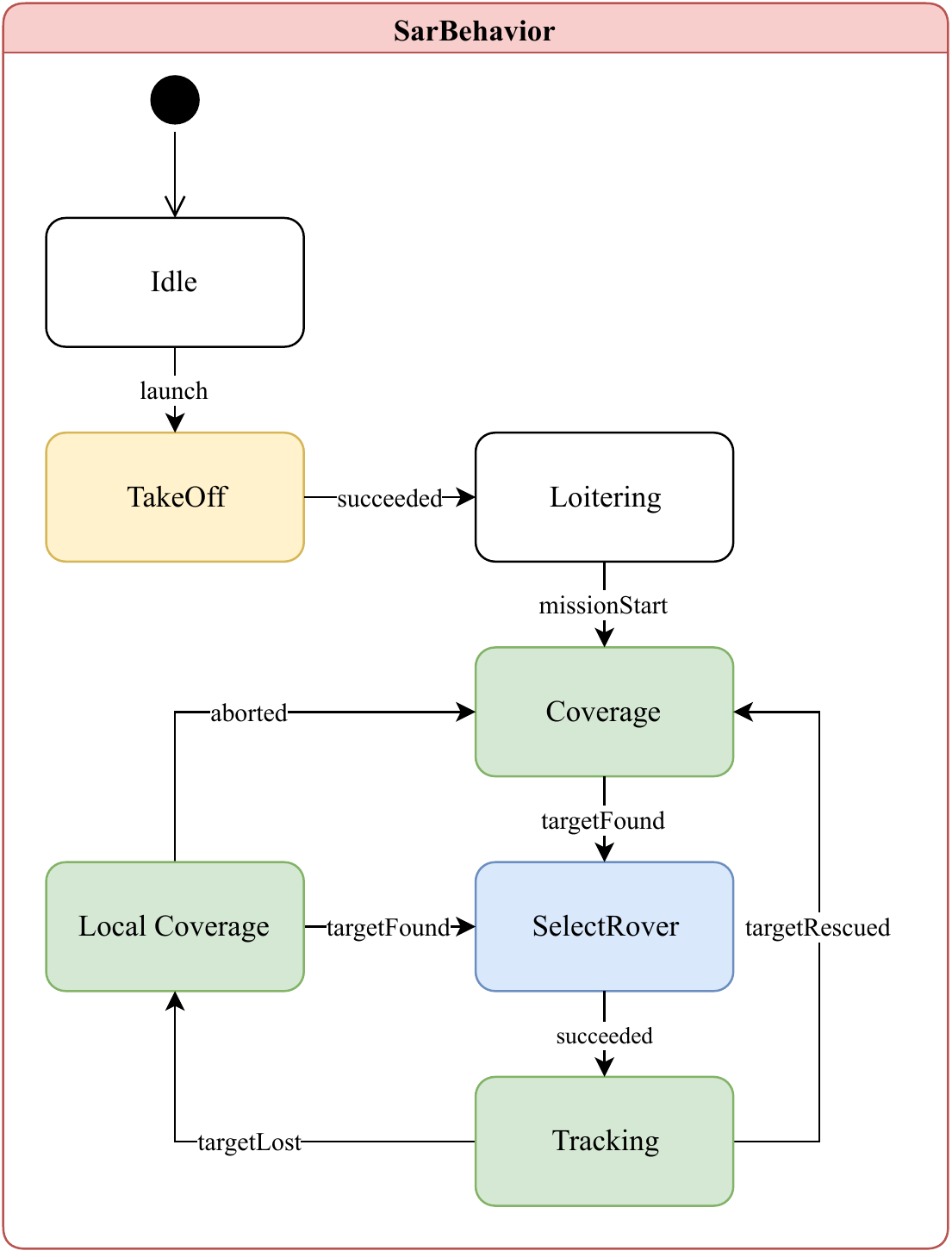}}%
\hfill
\subfloat[Legend]{%
\label{fig:fsm_legend}%
\includegraphics[width=0.5\linewidth]{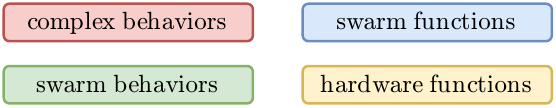}}%
\caption{The \ac{UAV} \ac{SAR} behavior \ac{FSM}.}
\label{fig:fsm}
\end{figure}

The first level in the hierarchy $L_1$, describes the set of parallel processes executed on each \ac{UAV} shown in Figure~\ref{fig:fsm_l1}. The actual \ac{SAR} behavior to accomplish the mission is modeled by the SarBehavior state. It is a complex behavior modeled as $L_2$ \ac{FSM}. The \textit{missionAbort} event allows to directly terminate the execution of the SarBehavior, including any behavior currently running in the corresponding $L_2$ \ac{FSM}. Hence, an event broadcast by the command and control station to all \acp{CPS} allows to completely stop the mission. There are several swarm functions and hardware functions running in parallel. They are detailed in the Tables~\ref{tab:swarm_functions} and \ref{tab:hardware_functions}, respectively.

The second level in the \ac{UAV} \ac{FSM} hierarchy $L_2$ describes the \ac{SAR} behavior executed by each \ac{UAV}. It is modeled by the \ac{FSM} shown in Figure~\ref{fig:fsm_l2}. After switching on a \ac{UAV}, it is in the Idle state. The event \textit{launch} emitted by the command and control station switches the \ac{UAV} to the TakeOff state where it lifts off. This function succeeds once the \ac{UAV} is at the desired altitude transitioning to the Loitering state in which the \ac{UAV} hovers stationary. The \ac{SAR} mission itself is started with the \textit{missionStart} event. The \ac{SAR} behavior starts with the Coverage state in which the \ac{UAV} searches for targets. If a target is found, the TargetMonitoring triggers the \textit{targetFound} event letting the \ac{UAV} switch to the SelectRover state. The \ac{UAV} communicates with all available \acp{UGV} to select the closest \ac{UGV} for moving to the target. While the selected \ac{UGV} moves towards the target, the \ac{UAV} performs the Tracking behavior to not lose the target until it is reached by the selected \ac{UGV}. It informs the \ac{UGV} about changes in position using the \textit{targetUpdate} event. If the \textit{targetRescued} event is received from the \ac{UGV}, the target is safe and the \ac{UAV} restarts the Coverage behavior to find other targets. If the target is lost before being rescued, the \ac{UAV} starts a LocalCoverage behavior. It allows the \ac{UAV} to circle around the last known target position to find the target again. If the target is found again, the \ac{UAV} again assigns the most suitable \ac{UGV}. Otherwise the \ac{UAV} restarts Coverage.

The \acp{FSM} use behaviors from the Swarm Library and the Abstraction Library. Tables~\ref{tab:swarm_behaviors} and \ref{tab:swarm_functions} list the swarm behaviors and functions used in the \ac{SAR} mission, respectively. The complete libraries are released as \ac{ROS} stacks\footnote{\url{https://wiki.ros.org/swarm_behaviors}}\footnote{\url{https://wiki.ros.org/swarm_functions}}. The hardware functions of the Abstraction Library used in the \ac{SAR} mission are given in Table~\ref{tab:hardware_functions}. There are further modules belonging to the different levels of the Abstraction Library released on GitHub. Please refer to the repositories of the Hardware Functions\footnote{\url{https://github.com/cpswarm/hardware_functions}} and Sensing and Actuation\footnote{\url{https://github.com/cpswarm/sensing_actuation}} libraries for further details. The events triggering the behavior state changes are summarized in Table~\ref{tab:events}.

\begin{table*}
\centering
\caption{The swarm functions used in the \ac{SAR} mission which are part of the Swarm Library.}
\begin{tabular}{m{2.2cm}|m{1.5cm}|m{1.5cm}|m{5.2cm}}
\textbf{Behavior} & \textbf{Input}                  & \textbf{Output}                                   & \textbf{Description} \\\hline\hline
TargetMonitoring  & camera footage                  & \acp{ID} of targets                               & Manages targets being detected by the swarm. It uses the on-board camera to detect targets and exchanges information about targets with the other \acp{CPS} in the swarm.\\\hline
SelectRover       & target \ac{ID}, target position & \ac{UGV} \ac{ID}, target \ac{ID}, target position & Assigns the closest idle \ac{UGV} for rescuing a target using the \textit{targetAssigned} event.\\\hline
\end{tabular}
\label{tab:swarm_functions}
\end{table*}

\begin{table*}
\centering
\caption{The hardware functions used in the \ac{SAR} mission which are part of the Abstraction Library.}
\begin{tabular}{m{2.2cm}|m{1.5cm}|m{1.5cm}|m{5.2cm}}
\textbf{Behavior} & \textbf{Input} & \textbf{Output} & \textbf{Description} \\\hline\hline
MissionAbort      & -              & -               & Lets a \ac{UAV} land.\\\hline
TakeOff           & altitude       & -               & Lets a \ac{UAV} lift off.\\\hline
\end{tabular}
\label{tab:hardware_functions}
\end{table*}

\begin{table*}
\centering
\caption{The swarm behaviors used in the \ac{SAR} mission which are part of the Swarm Library.}
\begin{tabular}{m{2.2cm}|m{1.5cm}|m{1.5cm}|m{5.2cm}}
\textbf{Behavior} & \textbf{Input}             & \textbf{Output}                 & \textbf{Description} \\\hline\hline
Coverage          & coverage area boundaries   & target \ac{ID}, target position & Lets a \ac{UAV} fly over the mission area in search of targets. It terminates once it finds a target.\\\hline
Tracking          & target \ac{ID}             & target position                 & Lets a \ac{UAV} keep track of a target that has been found. It informs other \acp{CPS} about position changes using the \textit{targetUpdate} event.\\\hline
LocalCoverage     & last known target position & target \ac{ID}, target position & Lets a \ac{UAV} search the local neighborhood for a lost target. This exploits the fact that a target has a high probability of being still close to the current position of the \ac{UAV}.\\\hline
\end{tabular}
\label{tab:swarm_behaviors}
\end{table*}

\begin{table}
\centering
\caption{Events used in the \ac{SAR} mission.}
\begin{tabular}{p{2.2cm}|p{3.45cm}|p{4cm}}
\textbf{Identifier} & \textbf{Data}                    & \textbf{Sender}             \\\hline\hline
launch              & -                                & command and control station \\\hline
missionStart        & -                                & command and control station \\\hline
missionAbort        & -                                & command and control station \\\hline
targetFound         & target \ac{ID}, target position  & swarm member                \\\hline
targetUpdate        & target \ac{ID}, target position  & swarm member                \\\hline
targetLost          & target \ac{ID}, target position  & swarm member                \\\hline
targetRescued       & target \ac{ID}                   & swarm member                \\\hline
targetAssigned      & target \ac{ID}, \ac{UGV} \ac{ID} & swarm member                \\\hline
\end{tabular}
\label{tab:events}
\end{table}

\section{Validation and Evaluation}
\label{sec:evaluation}

In this section we describe how the previously presented engineering process is validated and verified. Validation aids in creating a behavior model that reflects the desired swarm behavior while the verification process assures the generation of correct behavior code based on the modeled components. We focus on the models and the performance of the resulting implementations using the \ac{SAR} use case presented in the previous section. We deliberately omit the performance evaluation of every single workbench component taking part in the design process. We assume that during design time there are enough resources to effectively perform steps such as modeling, optimization, or code generation. The generated swarm behavior algorithms, however, must run efficiently on resource-limited hardware. Therefore, we evaluate the performance of the swarm algorithms resulting from the former steps to implicitly measure the effectiveness of the complete workbench. However, we acknowledge that there is the need to evaluate all parts of the workbench in greater detail, e.g., in terms of usability or degree of interoperability between the individual tools. The overall workbench performance could be measured through the speed up of \ac{CPS} design compared to a scenario without an integrated toolchain. Other performance indicators could be the quality of interaction and handovers between different users of the workbench and their satisfaction. However, this would require a completely new study which is out of the scope of this paper, which focuses only on the steps of modeling, code generation, and execution of the swarm algorithms.

\subsection{Validation and Verification Process}
In order to effectively perform the validation and verification of a swarm behavior, simulations are required. This is because the emerging behavior of the swarm cannot be predicted reliably using analytic methods.

As a first step, the models that have been created to describe the behavior need to be validated in order to prove that they exhibit the desired behavior of the swarm. As a second step, the code generated from the hardware and behavior models needs to be verified in order to prove that it exhibits the same behavior.

For effective validation and verification, contradicting requirements exist. On the one hand, the validation process should report back any flaws of the model as quickly as possible to avoid a waste of effort while implementing the model. This means that the implementation effort should be low and the run time of the simulation should be short. On the other hand, to get a meaningful verification, the resulting data should be as accurate as possible. This means that the simulation should be as realistic as possible to minimize the problem of the reality gap, i.e., transferability of the generated code from simulation to the real world. These contradicting requirements cannot easily be reconciled using a single simulation, even though some efforts have been done in this direction~\cite{Pinciroli12}. A common approach is to use a hierarchy of different simulations and real-world experiments adding increasingly more detail~\cite{Faigl2015,Rand2007}. An overview of the different levels of realism used during our validation and verification process can be seen in Table~\ref{tab:properties-realism}.
\begin{table*}
\centering
\caption{Properties of levels of realism and metrics for evaluation.}
\label{tab:properties-realism}
\begin{tabular}{r||c|c|c}
                      & \textbf{Abstract}   & \textbf{Detailed}   & \textbf{Real}    \\
                      & \textbf{Simulation} & \textbf{Simulation} & \textbf{World}   \\\hline\hline
\textbf{Metrics}      &                     &                     &                  \\
Implementation effort & low                 & high                & very high        \\
Run time              & short               & long                & real time        \\
Accuracy of results   & low                 & high                & exact            \\\hline
\textbf{Properties}   &                     &                     &                  \\
World model           & abstract            & detailed            & realistic        \\
Hardware model        & abstract            & detailed            & realistic        \\
Behavior model        & detailed            & realistic           & realistic        \\
Agent communication   & shared memory       & network socket      & wireless network \\
Time                  & discrete            & continuous          & continuous       \\
Space                 & discrete            & continuous          & continuous
\end{tabular}
\end{table*}
The table classifies each level w.r.t. the metrics implementation effort, run time, and accuracy of results.

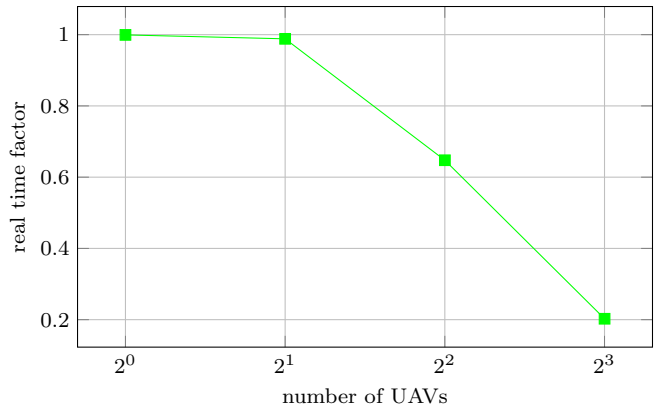
\begin{figure}
\centering
\tikzsetnextfilename{real_time_factor}
\begin{tikzpicture}

\begin{semilogxaxis}[
width=0.75\linewidth,
height=0.5\linewidth,
xlabel={number of \acp{UAV}},
ylabel={real time factor},,
log basis x=2,
grid=both,
font=\footnotesize,
legend style={at={(1,0.5)}, anchor=east, font=\scriptsize},
legend cell align=left,
/pgf/number format/1000 sep={}
]

\addplot[green,mark=square*]
table[x=uav,y=rtf] {data/real_time_factor/real_time_factor_ros.txt};

\end{semilogxaxis}

\end{tikzpicture}
\caption{\ac{RTF} of the detailed coverage simulation over number of \acp{UAV}.}
\label{fig:real_time_factor}
\end{figure}

In order to demonstrate the need for the multi-leveled evaluation, we record the \ac{RTF}
\begin{equation}
\gamma = \frac{t_{\rm sim}}{t_{\rm wall}}
\end{equation}
which relates the simulated time $t_{\rm sim}$ to the actual computing time $t_{\rm wall}$. While the former measures the time experienced by the \acp{CPS} in the simulation, the latter measures the real, physical time experienced by the human experimenter. Figure~\ref{fig:real_time_factor} shows the real time factor for the detailed simulation. The abstract simulation is not included as the high level of abstraction prevents an exact mapping of the abstract time ticks that simulate seconds. The results naturally depend on the performance of the computer used for the simulations. For our measurements, we used a laptop computer with an Intel Core i7-5500U \ac{CPU}\footnote{\url{https://ark.intel.com/content/www/us/en/ark/products/85214/intel-core-i7-5500u-processor-4m-cache-up-to-3-00-ghz.html}} with two cores at \SI{2.4}{GHz} providing four threads and \SI{8}{\giga\byte} of memory.
The results show that the \ac{RTF} is always below one, meaning that the simulation never speeds up the experiments compared to real-world experiments. Furthermore, the performance significantly drops once the number of \acp{UAV} exceeds 2. This suggests that the detailed simulation requires one \ac{CPU} core for each simulated \ac{CPS}. If this condition is satisfied, the simulation runs nearly in real time.

We can compare the detailed and the abstract simulations through the total simulation time for all experiments. The abstract simulation takes about \SI{4}{\minute} and \SI{40}{\second} to complete a total of 200 simulation runs. The detailed simulation takes about \SI{27}{\day}, \SI{11}{\hour} and \SI{51}{\minute} to complete a total of 246 simulation runs. The difference of several orders of magnitude shows the importance for the abstract simulation during the engineering process.

\subsection{Performance Evaluation}
The behavior of the swarm of \acp{CPS} that is designed using the process described in Section~\ref{sec:method} is evaluated experimentally through simulations. As described in the previous section, we use two different types of simulations to measure the performance of the swarm behavior: first, an abstract simulation that provides results fast but less accurately. By observing the behavior of the simulated \acp{CPS}, it is possible to validate that the design process yields the desired results. The performance measures allow to determine whether the emerging swarm behavior meets the requirements. Second, a detailed simulation that provides more realistic results but requires more simulation time. The simulations provide results that allow measuring the performance of the emergent swarm behavior using well defined performance metrics. We qualitatively compare the outcomes of the two simulations in order to verify the outcomes of the design process. The abstract simulation evaluates the complete \ac{SAR} behavior consisting of coverage, tracking, and reaching the target. The latter will be referred to as rescuing in the following. Because of the high implementation and simulation effort of the detailed simulation, only the coverage behavior is evaluated throughout this paper.

\subsubsection{Coverage}

Throughout this paper, we evaluate the coverage behavior using both, the abstract and the detailed simulation. It is implemented as a random direction algorithm where the \acp{UAV} fly on a straight line until they encounter an obstacle or the area boundary, and then randomly change their direction. While this algorithm is well suited for bounded environments, unbounded environments would require a different algorithm such as random walk~\citep{Pearson1905}.The results are then compared qualitatively to verify the outcomes of the design process. We evaluate the coverage efficiency by measuring the time it takes to cover a given percentage $p$ of the environment area. We chose the following percentages of interest
\begin{equation}
p \in \{\SI{50}{\percent},\SI{75}{\percent},\SI{87.5}{\percent},\SI{93.75}{\percent},\SI{96.875}{\percent},\SI{100}{\percent}\}
\label{eq:percentages_of_interest}
\end{equation}
where the percentages are more and more fine grained the higher they get. Percentages below \SI{50}{\percent} are not considered to be effective enough for the scenario. The environment in which the \ac{SAR} takes place is a bounded, rectangular region without any obstacles. Unbounded environments would work just as well, however the performance metric would need to measure the covered area in absolute terms. All \acp{CPS} are placed in the center of the south side of the environment with equal spacing. Once the mission starts, all \acp{UAV} take off and start covering the area by departing in diverging directions chosen by dividing the available \SI{180}{\degree} regularly by the number of \acp{UAV}. The simulation is performed with increasing swarm size using one, two, four, and eight \acp{UAV}.

\subsubsection{Coverage}

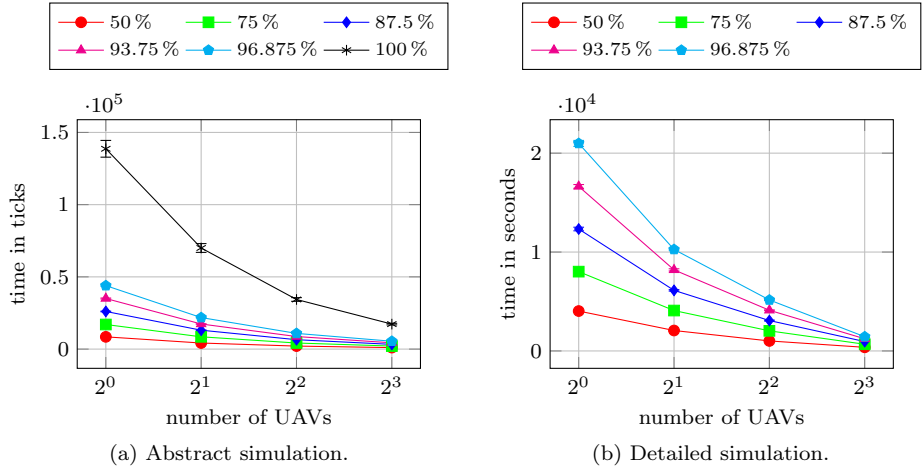
\begin{figure*}
\centering
\tikzsetnextfilename{coverage_time_abstract}
\subfloat[Abstract simulation.]{%
\label{fig:coverage_time_abstract}%
%% Graph for Coverage / Netlogo

\begin{tikzpicture}

\begin{semilogxaxis}[
width=0.5\linewidth,
height=0.4\linewidth,
xlabel={number of \acp{UAV}},
ylabel={time in \si{ticks}},
log basis x=2,
grid=both,
font=\footnotesize,
legend style={at={(0.5,1.2)}, anchor=south, font=\scriptsize, legend columns=3},
legend cell align=left,
/pgf/number format/1000 sep={}
]

\addplot[red,mark=*,error bars/y dir=both,error bars/y explicit]
table[x=uav,y=time,y error=ci] {data/sim_netlogo/timeNL_50.txt};
\addlegendentry{ \SI{50}{\percent} }

\addplot[green,mark=square*,error bars/y dir=both,error bars/y explicit]
table[x=uav,y=time,y error=ci] {data/sim_netlogo/timeNL_75.txt};
\addlegendentry{ \SI{75}{\percent} }

\addplot[blue,mark=diamond*,error bars/y dir=both,error bars/y explicit]
table[x=uav,y=time,y error=ci] {data/sim_netlogo/timeNL_88.txt};
\addlegendentry{ \SI{87.5}{\percent} }

\addplot[magenta,mark=triangle*,error bars/y dir=both,error bars/y explicit]
table[x=uav,y=time,y error=ci] {data/sim_netlogo/timeNL_94.txt};
\addlegendentry{ \SI{93.75}{\percent} }

\addplot[cyan,mark=pentagon*,error bars/y dir=both,error bars/y explicit]
table[x=uav,y=time,y error=ci] {data/sim_netlogo/timeNL_97.txt};
\addlegendentry{ \SI{96.875}{\percent} }

\addplot[black,mark=asterisk,error bars/y dir=both,error bars/y explicit]
table[x=uav,y=time,y error=ci] {data/sim_netlogo/timeNL_100.txt};
\addlegendentry{ \SI{100}{\percent} }

\end{semilogxaxis}

\end{tikzpicture}}%
\hfill
\tikzsetnextfilename{coverage_time_detailed}
\subfloat[Detailed simulation.]{%
\label{fig:coverage_time_detailed}%
\begin{tikzpicture}

\begin{semilogxaxis}[
width=0.5\linewidth,
height=0.4\linewidth,
xlabel={number of \acp{UAV}},
ylabel={time in \si{seconds}},
log basis x=2,
grid=both,
font=\footnotesize,
legend style={at={(0.5,1.2)}, anchor=south, font=\scriptsize, legend columns=3},
legend cell align=left,
/pgf/number format/1000 sep={}
]

\addplot[red,mark=*,error bars/y dir=both,error bars/y explicit]
table[x=uav,y=time,y error=ci] {data/sim_ros/time_50.txt};
\addlegendentry{ \SI{50}{\percent} }

\addplot[green,mark=square*,error bars/y dir=both,error bars/y explicit]
table[x=uav,y=time,y error=ci] {data/sim_ros/time_75.txt};
\addlegendentry{ \SI{75}{\percent} }

\addplot[blue,mark=diamond*,error bars/y dir=both,error bars/y explicit]
table[x=uav,y=time,y error=ci] {data/sim_ros/time_88.txt};
\addlegendentry{ \SI{87.5}{\percent} }

\addplot[magenta,mark=triangle*,error bars/y dir=both,error bars/y explicit]
table[x=uav,y=time,y error=ci] {data/sim_ros/time_94.txt};
\addlegendentry{ \SI{93.75}{\percent} }

\addplot[cyan,mark=pentagon*,error bars/y dir=both,error bars/y explicit]
table[x=uav,y=time,y error=ci] {data/sim_ros/time_97.txt};
\addlegendentry{ \SI{96.875}{\percent} }

%\addplot[black,mark=asterisk,error bars/y dir=both,error bars/y explicit]
%table[x=uav,y=time,y error=ci] {data/sim_ros/time_100.txt};
%\addlegendentry{ \SI{100}{\percent} }

\end{semilogxaxis}

\end{tikzpicture}}%
\caption{Simulation coverage time over number of \acp{UAV} for varying coverage percentage rates.}
\label{fig:coverage_time}
\end{figure*}

\paragraph{The abstract simulation} is implemented in Netlogo~6.1.1~\cite{Wilensky1999, Wilensky2015}, an agent-based simulator using discrete time and space. The world is modeled as a square of $201 \times 201$ patches. The \acp{UAV} are placed four patches apart and fly with a constant speed of one patch per tick. Their \ac{FOV} consists of the patch they are currently above plus the surrounding eight patches, i.e., the Moore neighborhood. The simulation is repeated 50 times per configuration. The results visualized in Figure~\ref{fig:coverage_time_abstract} show the coverage efficiency, i.e., the average time for covering the different percentages of the environment, with error bars showing the \SI{95}{\percent} confidence interval.
The coverage time is inversely proportional to the swarm size. It scales very well with increasing swarm size as doubling the number of \acp{UAV} approximately bisects the required coverage time. Increasing the coverage percentage linearly increases the coverage time exponentially. This is because the uncovered space decreases over time, increasing the redundancy of the coverage. This becomes evident for complete coverage (i.e. \SI{100}{\percent}) where the coverage time increases dramatically. This is because the random direction algorithm makes it very unlikely to choose exactly the right direction towards the last unexplored areas at the end of the exploration.

\paragraph{The detailed simulation} is based on \ac{ROS} using the Gazebo simulator~\cite{Koenig2004} which allows to perform high fidelity simulations based on a physics engine that simulates continuous space and time. As the simulations build on \ac{ROS} the same behavior code generated for execution on the \acp{CPS} using the swarm and abstraction libraries can be used in simulation. This allows effective verification of the generated code using many simulation runs before the actual hardware deployment takes place. The random direction algorithm used for these simulations is released as \ac{ROS} package\footnote{\url{https://wiki.ros.org/uav_random_direction}}. It is initialized with the \acp{UAV} being placed equally spaced with \SI{2}{\meter} separation. The altitude of operation is \SI{1.5}{\meter} above the ground giving the \acp{UAV} a \ac{FOV} radius of approximately \SI{0.5}{\meter}. They fly with an average speed of \SI{0.8}{\meter\per\second}. The coverage is performed in an area of size $\SI{66}{\meter} \times \SI{66}{\meter}$. To get reliable results, the simulation is repeated multiple times for each parameter setting in order to reach a relative error of the data sample below \SI{10}{\percent} with a confidence of \SI{99}{\percent}. The results in terms of coverage efficiency can be seen in Figure~\ref{fig:coverage_time_detailed}. The results are very similar to the ones from the abstract simulation. The coverage time is inversely proportional to the swarm size and increases exponentially with increasing coverage percentage. The absolute values are not directly comparable between the simulations as the difference in abstraction is too large. Nevertheless, if we assume that a simulated tick in abstract simulation takes about \SI{0.5}{\second}, there is a relative difference of approximately \SI{5}{\percent} in coverage time between levels of abstraction. In the detailed simulation, it was unfeasible to reach \SI{100}{\percent} coverage due to simulation time constraints. Therefore, results are shown only for a maximum value of \SI{96.875}{\percent}. These results verify that the behavior obtained from the generated code meets the modeled behavior.

\subsubsection{Search and Rescue}
\begin{figure*}
\centering
\tikzsetnextfilename{sar_time_abstract_1}
\subfloat[1 \ac{UAV}.]{%% Graph for SAR / Netlogo with  1  drone

\begin{tikzpicture}

\begin{semilogxaxis}[
width=0.5\textwidth,
height=0.4\textwidth,
xlabel={number of targets},
ylabel={time in \si{ticks}},
log basis x=2,
grid=both,
font=\footnotesize,
legend style={at={(0.5,1.2)}, anchor=south, font=\scriptsize, legend columns=3},
legend cell align=left,
/pgf/number format/1000 sep={}
]

%-- 1st
\addplot[brown,mark=triangle*,error bars/y dir=both,error bars/y explicit]
table[x=victims,y=timeF,y error=ciR] {data/sim_netlogo/timeNL_SAR_1_1.txt};
\addlegendentry{ \SI{1}st }

\addplot[brown,dashed,mark=triangle*,error bars/y dir=both,error bars/y explicit]
table[x=victims,y=timeR,y error=ciF, forget plot] {data/sim_netlogo/timeNL_SAR_1_1.txt};

%-- 50%
\addplot[red,mark=*,error bars/y dir=both,error bars/y explicit]
table[x=victims,y=timeF,y error=ciR] {data/sim_netlogo/timeNL_SAR_1_50.txt};
\addlegendentry{ \SI{50}{\percent} }

\addplot[red,dashed,mark=*,error bars/y dir=both,error bars/y explicit]
table[x=victims,y=timeR,y error=ciF, forget plot] {data/sim_netlogo/timeNL_SAR_1_50.txt};

%-- 75%
\addplot[green,mark=square*,error bars/y dir=both,error bars/y explicit]
table[x=victims,y=timeF,y error=ciF] {data/sim_netlogo/timeNL_SAR_1_75.txt};
\addlegendentry{ \SI{75}{\percent} }

\addplot[green,dashed,mark=square*,error bars/y dir=both,error bars/y explicit]
table[x=victims,y=timeR,y error=ciR, forget plot] {data/sim_netlogo/timeNL_SAR_1_75.txt};

%-- 90%
\addplot[blue,mark=diamond*,error bars/y dir=both,error bars/y explicit]
table[x=victims,y=timeF,y error=ciF] {data/sim_netlogo/timeNL_SAR_1_90.txt};
\addlegendentry{ \SI{90}{\percent} }

\addplot[blue,dashed,mark=diamond*,error bars/y dir=both,error bars/y explicit]
table[x=victims,y=timeR,y error=ciR, forget plot] {data/sim_netlogo/timeNL_SAR_1_90.txt};

%-- 100%
\addplot[black,mark=asterisk,error bars/y dir=both,error bars/y explicit]
table[x=victims,y=timeF,y error=ciF] {data/sim_netlogo/timeNL_SAR_1_100.txt};
\addlegendentry{ \SI{100}{\percent} }

\addplot[black,dashed,mark=asterisk,error bars/y dir=both,error bars/y explicit]
table[x=victims,y=timeR,y error=ciR, forget plot] {data/sim_netlogo/timeNL_SAR_1_100.txt};

\end{semilogxaxis}

\end{tikzpicture}}%
\hfill
\tikzsetnextfilename{sar_time_abstract_2}
\subfloat[2 \acp{UAV}.]{%% Graph for SAR / Netlogo with  2  drones

\begin{tikzpicture}

\begin{semilogxaxis}[
width=0.5\textwidth,
height=0.4\textwidth,
xlabel={number of targets},
ylabel={time in \si{ticks}},
log basis x=2,
grid=both,
font=\footnotesize,
legend style={at={(0.5,1.2)}, anchor=south, font=\scriptsize, legend columns=3},
legend cell align=left,
/pgf/number format/1000 sep={}
]

%-- 1st
\addplot[brown,mark=triangle*,error bars/y dir=both,error bars/y explicit]
table[x=victims,y=timeF,y error=ciR] {data/sim_netlogo/timeNL_SAR_2_1.txt};
\addlegendentry{ \SI{1}st }

\addplot[brown,dashed,mark=triangle*,error bars/y dir=both,error bars/y explicit]
table[x=victims,y=timeR,y error=ciF, forget plot] {data/sim_netlogo/timeNL_SAR_2_1.txt};

%-- 50%
\addplot[red,mark=*,error bars/y dir=both,error bars/y explicit]
table[x=victims,y=timeF,y error=ciR] {data/sim_netlogo/timeNL_SAR_2_50.txt};
\addlegendentry{ \SI{50}{\percent} }

\addplot[red,dashed,mark=*,error bars/y dir=both,error bars/y explicit]
table[x=victims,y=timeR,y error=ciF, forget plot] {data/sim_netlogo/timeNL_SAR_2_50.txt};

%-- 75%
\addplot[green,mark=square*,error bars/y dir=both,error bars/y explicit]
table[x=victims,y=timeF,y error=ciF] {data/sim_netlogo/timeNL_SAR_2_75.txt};
\addlegendentry{ \SI{75}{\percent} }

\addplot[green,dashed,mark=square*,error bars/y dir=both,error bars/y explicit]
table[x=victims,y=timeR,y error=ciR, forget plot] {data/sim_netlogo/timeNL_SAR_2_75.txt};

%-- 90%
\addplot[blue,mark=diamond*,error bars/y dir=both,error bars/y explicit]
table[x=victims,y=timeF,y error=ciF] {data/sim_netlogo/timeNL_SAR_2_90.txt};
\addlegendentry{ \SI{90}{\percent} }

\addplot[blue,dashed,mark=diamond*,error bars/y dir=both,error bars/y explicit]
table[x=victims,y=timeR,y error=ciR, forget plot] {data/sim_netlogo/timeNL_SAR_2_90.txt};

%-- 100%
\addplot[black,mark=asterisk,error bars/y dir=both,error bars/y explicit]
table[x=victims,y=timeF,y error=ciF] {data/sim_netlogo/timeNL_SAR_2_100.txt};
\addlegendentry{ \SI{100}{\percent} }

\addplot[black,dashed,mark=asterisk,error bars/y dir=both,error bars/y explicit]
table[x=victims,y=timeR,y error=ciR, forget plot] {data/sim_netlogo/timeNL_SAR_2_100.txt};

\end{semilogxaxis}

\end{tikzpicture}}%
\\
\tikzsetnextfilename{sar_time_abstract_4}
\subfloat[4 \ac{UAV}.]{%% Graph for SAR / Netlogo with  4  drones

\begin{tikzpicture}

\begin{semilogxaxis}[
width=0.5\textwidth,
height=0.4\textwidth,
xlabel={number of targets},
ylabel={time in \si{ticks}},
log basis x=2,
grid=both,
font=\footnotesize,
legend style={at={(0.5,1.2)}, anchor=south, font=\scriptsize, legend columns=3},
legend cell align=left,
/pgf/number format/1000 sep={}
]

%-- 1st
\addplot[brown,mark=triangle*,error bars/y dir=both,error bars/y explicit]
table[x=victims,y=timeF,y error=ciR] {data/sim_netlogo/timeNL_SAR_4_1.txt};
\addlegendentry{ \SI{1}st }

\addplot[brown,dashed,mark=triangle*,error bars/y dir=both,error bars/y explicit]
table[x=victims,y=timeR,y error=ciF, forget plot] {data/sim_netlogo/timeNL_SAR_4_1.txt};

%-- 50%
\addplot[red,mark=*,error bars/y dir=both,error bars/y explicit]
table[x=victims,y=timeF,y error=ciR] {data/sim_netlogo/timeNL_SAR_4_50.txt};
\addlegendentry{ \SI{50}{\percent} }

\addplot[red,dashed,mark=*,error bars/y dir=both,error bars/y explicit]
table[x=victims,y=timeR,y error=ciF, forget plot] {data/sim_netlogo/timeNL_SAR_4_50.txt};

%-- 75%
\addplot[green,mark=square*,error bars/y dir=both,error bars/y explicit]
table[x=victims,y=timeF,y error=ciF] {data/sim_netlogo/timeNL_SAR_4_75.txt};
\addlegendentry{ \SI{75}{\percent} }

\addplot[green,dashed,mark=square*,error bars/y dir=both,error bars/y explicit]
table[x=victims,y=timeR,y error=ciR, forget plot] {data/sim_netlogo/timeNL_SAR_4_75.txt};

%-- 90%
\addplot[blue,mark=diamond*,error bars/y dir=both,error bars/y explicit]
table[x=victims,y=timeF,y error=ciF] {data/sim_netlogo/timeNL_SAR_4_90.txt};
\addlegendentry{ \SI{90}{\percent} }

\addplot[blue,dashed,mark=diamond*,error bars/y dir=both,error bars/y explicit]
table[x=victims,y=timeR,y error=ciR, forget plot] {data/sim_netlogo/timeNL_SAR_4_90.txt};

%-- 100%
\addplot[black,mark=asterisk,error bars/y dir=both,error bars/y explicit]
table[x=victims,y=timeF,y error=ciF] {data/sim_netlogo/timeNL_SAR_4_100.txt};
\addlegendentry{ \SI{100}{\percent} }

\addplot[black,dashed,mark=asterisk,error bars/y dir=both,error bars/y explicit]
table[x=victims,y=timeR,y error=ciR, forget plot] {data/sim_netlogo/timeNL_SAR_4_100.txt};

\end{semilogxaxis}

\end{tikzpicture}}%
\hfill
\tikzsetnextfilename{sar_time_abstract_8}
\subfloat[8 \acp{UAV}.]{%% Graph for SAR / Netlogo with  8  drones

\begin{tikzpicture}

\begin{semilogxaxis}[
width=0.5\textwidth,
height=0.4\textwidth,
xlabel={number of targets},
ylabel={time in \si{ticks}},
log basis x=2,
grid=both,
font=\footnotesize,
legend style={at={(0.5,1.2)}, anchor=south, font=\scriptsize, legend columns=3},
legend cell align=left,
/pgf/number format/1000 sep={}
]

%-- 1st
\addplot[brown,mark=triangle*,error bars/y dir=both,error bars/y explicit]
table[x=victims,y=timeF,y error=ciR] {data/sim_netlogo/timeNL_SAR_8_1.txt};
\addlegendentry{ \SI{1}st }

\addplot[brown,dashed,mark=triangle*,error bars/y dir=both,error bars/y explicit]
table[x=victims,y=timeR,y error=ciF, forget plot] {data/sim_netlogo/timeNL_SAR_8_1.txt};

%-- 50%
\addplot[red,mark=*,error bars/y dir=both,error bars/y explicit]
table[x=victims,y=timeF,y error=ciR] {data/sim_netlogo/timeNL_SAR_8_50.txt};
\addlegendentry{ \SI{50}{\percent} }

\addplot[red,dashed,mark=*,error bars/y dir=both,error bars/y explicit]
table[x=victims,y=timeR,y error=ciF, forget plot] {data/sim_netlogo/timeNL_SAR_8_50.txt};

%-- 75%
\addplot[green,mark=square*,error bars/y dir=both,error bars/y explicit]
table[x=victims,y=timeF,y error=ciF] {data/sim_netlogo/timeNL_SAR_8_75.txt};
\addlegendentry{ \SI{75}{\percent} }

\addplot[green,dashed,mark=square*,error bars/y dir=both,error bars/y explicit]
table[x=victims,y=timeR,y error=ciR, forget plot] {data/sim_netlogo/timeNL_SAR_8_75.txt};

%-- 90%
\addplot[blue,mark=diamond*,error bars/y dir=both,error bars/y explicit]
table[x=victims,y=timeF,y error=ciF] {data/sim_netlogo/timeNL_SAR_8_90.txt};
\addlegendentry{ \SI{90}{\percent} }

\addplot[blue,dashed,mark=diamond*,error bars/y dir=both,error bars/y explicit]
table[x=victims,y=timeR,y error=ciR, forget plot] {data/sim_netlogo/timeNL_SAR_8_90.txt};

%-- 100%
\addplot[black,mark=asterisk,error bars/y dir=both,error bars/y explicit]
table[x=victims,y=timeF,y error=ciF] {data/sim_netlogo/timeNL_SAR_8_100.txt};
\addlegendentry{ \SI{100}{\percent} }

\addplot[black,dashed,mark=asterisk,error bars/y dir=both,error bars/y explicit]
table[x=victims,y=timeR,y error=ciR, forget plot] {data/sim_netlogo/timeNL_SAR_8_100.txt};

\end{semilogxaxis}

\end{tikzpicture}}%
\caption{Abstract simulation rescue time over number of targets. Continuous lines represent average times for targets to be found by a \ac{UAV}, dashed lines show average times until targets are reached by a \ac{UGV}.}
\label{fig:sar_time_abstract}
\end{figure*}
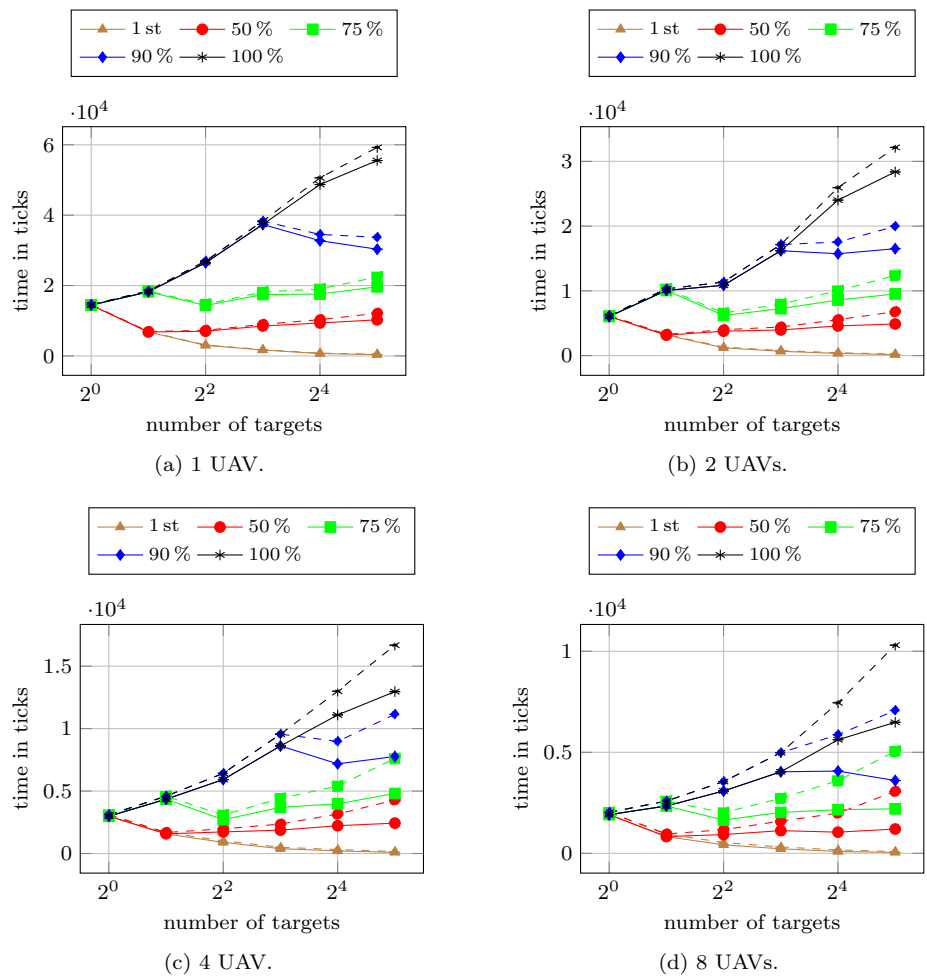

The abstract simulation for the \ac{SAR} scenario consists of an extension of the coverage scenario simulation described above. We extended the simulation by adding targets to be rescued and \acp{UGV} performing the rescue. Targets are distributed uniform randomly in the world. A target is considered found when it comes into the \ac{FOV} of a \ac{UAV}; this is the search time for this target. The \ac{UAV} then hovers over the site until a \ac{UGV} reaches the target. To avoid \acp{UAV} from being blocked in tracking, there is one \ac{UGV} exclusively assigned to each \ac{UAV}. The \ac{UGV} starts from the same initial location as the \ac{UAV} that found the target and drives towards the target in a straight line with a constant speed of one patch per tick. As soon as the \ac{UGV} reaches the target, this target is considered to be rescued and the \ac{UAV} continues to search for further targets. We perform simulations to measure the rescue efficiency, i.e., the time until all targets are rescued. We perform the simulation with one, two, four, eight, 16, and 32 targets. Each configuration is simulated with 50 repetitions for statistical significance. Figure~\ref{fig:sar_time_abstract} shows the results of the abstract simulations for the \ac{SAR} scenario.
The figures show the average search times of targets (continuous lines) and the average rescue times of targets (dashed lines) for the first target, \SI{50}{\percent}, \SI{75}{\percent}, \SI{90}{\percent}, and \SI{100}{\percent} of the targets. As for the coverage simulation, the search times are inversely proportional to the number of \acp{UAV}. The rescue times are naturally larger than the search times since they include the traveling time of the \acp{UGV}. This traveling time stays constant, regardless of the number of employed \acp{UAV}. Therefore, with increasing number of \acp{UAV} and decreasing search time, the rescue time relative to the search time becomes larger. As expected, an increasing number of targets leads to increased search and rescue times. Exceptions only occur for low percentages of targets, e.g., finding one out of two targets is naturally faster than finding a single target. It should furthermore be noted that the target percentages are always rounded up since the number of targets is a discrete quantity. This leads to the effect where an increased number of targets decreases the search and rescue times. For example, looking at the \SI{75}{\percent} line, one can observe finding two out of two targets is more costly than finding three out of four targets. These results confirm that the \ac{SAR} mission performs as expected and hence validate the modeled behavior.

\subsection{Proof of Concept}
\begin{figure*}
\centering
\includegraphics[width=\linewidth]{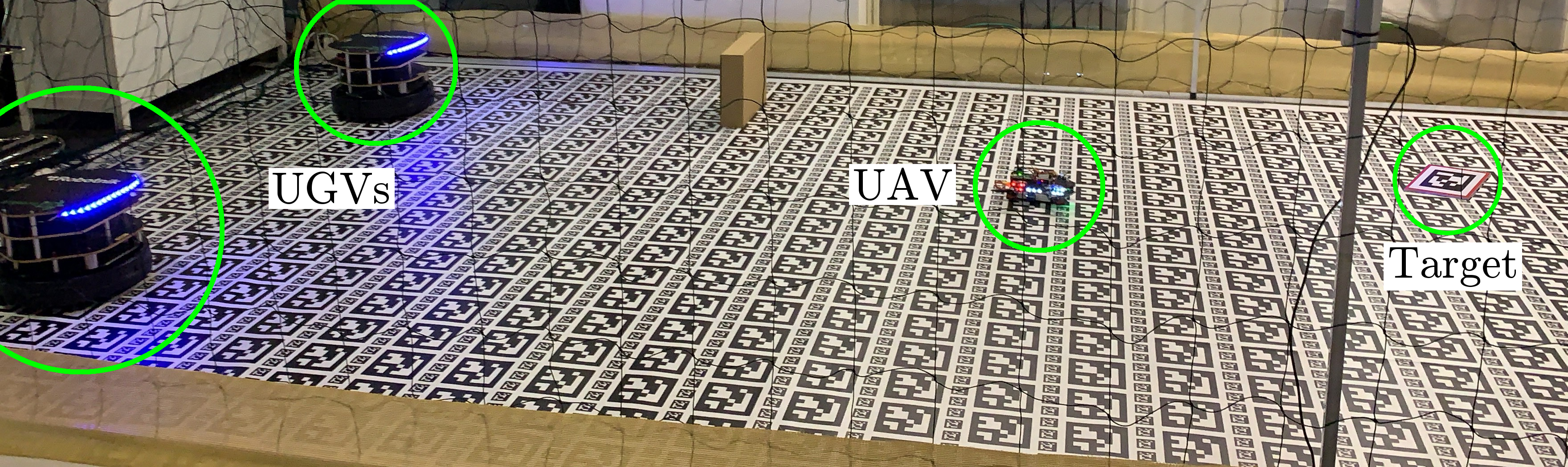}
\caption{The arena for the \ac{SAR} demonstration containing \acp{UGV}, a \ac{UAV}, and target.}
\label{fig:arena}
\end{figure*}

\begin{figure}
\centering
\tikzsetnextfilename{event_time_demo}
\begin{tikzpicture}

\begin{axis}[
width=0.75\linewidth,
height=0.5\linewidth,
xlabel={time in \si{seconds}},
xmajorgrids,
ytick={1,2,3,4,5},
yticklabels={launch, missionStart, targetFound, targetAssigned, targetRescued},
]

%                       event: [lw, lq, m, uq, uw, [outliers]]
%            cmd/takeoff/goal: [0.1, 0.4, 0.56, 0.77, 1.04, []] (21 runs)
%          cmd/takeoff/result: [12.21, 12.91, 13.1, 13.4, 13.4, [14.91, 14.6, 15.17, 14.5, 14.73]] (21 runs)
% bridge/events/mission_start: [13.93, 16.47, 17.77, 20.5, 24.0, [30.0, 31.15]] (21 runs)
%           uav_coverage/goal: [14.17, 17.44, 18.63, 20.84, 24.17, [30.98, 31.41]] (21 runs)
% bridge/events/cps_selection: [51.12, 73.88, 83.25, 89.4, 99.41, [115.97]] (21 runs)
%         uav_coverage/result: [80.72, 90.2, 96.58, 104.08, 111.82, [128.25, 58.81]] (21 runs)
%        cmd/assign_task/goal: [80.74, 90.21, 96.61, 104.1, 111.84, [128.26, 58.83]] (21 runs)
%                cps_selected: [82.74, 92.22, 98.61, 106.1, 113.84, [130.27, 60.83]] (21 runs)
%      cmd/assign_task/result: [82.75, 92.22, 98.61, 106.1, 113.84, [130.27, 60.83]] (21 runs)
%           uav_tracking/goal: [82.76, 92.23, 98.62, 106.11, 113.85, [130.28, 60.85]] (21 runs)
%                 target_lost: [85.0, 95.58, 100.87, 108.58, 116.48, [132.69, 63.14]] (21 runs)
%                target_found: [89.17, 95.75, 103.49, 106.74, 113.82, [126.28]] (9 runs)
% bridge/events/mission_abort: [147.81, 155.4, 166.34, 176.94, 203.28, [121.97]] (19 runs)
%         uav_tracking/cancel: [147.85, 155.44, 166.38, 176.98, 203.31, [122.0]] (19 runs)
%         uav_tracking/result: [149.64, 155.65, 166.98, 177.12, 205.1, [122.74]] (19 runs)

% cmd/takeoff/goal
\addplot+[
boxplot prepared={
lower whisker=0.1,
lower quartile=0.4,
median=0.56,
upper quartile=0.77,
upper whisker=1.04
},
]
coordinates {};

% bridge/events/mission_start
\addplot+[
boxplot prepared={
lower whisker=13.93,
lower quartile=16.47,
median=17.77,
upper quartile=20.5,
upper whisker=24.0
},
]
coordinates {(0,30.0) (0,31.15)};

% bridge/events/cps_selection
\addplot+[
boxplot prepared={
lower whisker=51.12,
lower quartile=73.88,
median=83.25,
upper quartile=89.4,
upper whisker=99.41
},
]
coordinates {(0,115.97)};

% uav_coverage/result
% \addplot+[
% boxplot prepared={
% lower whisker=80.72,
% lower quartile=90.2,
% median=96.58,
% upper quartile=104.08,
% upper whisker=111.82
% },
% ]
% coordinates {(0,128.25) (0,58.81)};

% cps_selected
\addplot+[
boxplot prepared={
lower whisker=82.74,
lower quartile=92.22,
median=98.61,
upper quartile=106.1,
upper whisker=113.84
},
]
coordinates {(0,130.27) (0,60.83)};

% bridge/events/mission_abort
\addplot+[
boxplot prepared={
lower whisker=147.81,
lower quartile=155.4,
median=166.34,
upper quartile=176.94,
upper whisker=203.28
},
]
coordinates {(0,121.97)};

% \addplot+[
% boxplot prepared={
% lower whisker=,
% lower quartile=,
% median=,
% upper quartile=,
% upper whisker=
% },
% ]
% coordinates {(0,) (0,)};

\end{axis}

\end{tikzpicture}
\caption{Timeline of demonstration.}
\label{fig:event_time_demo}
\end{figure}
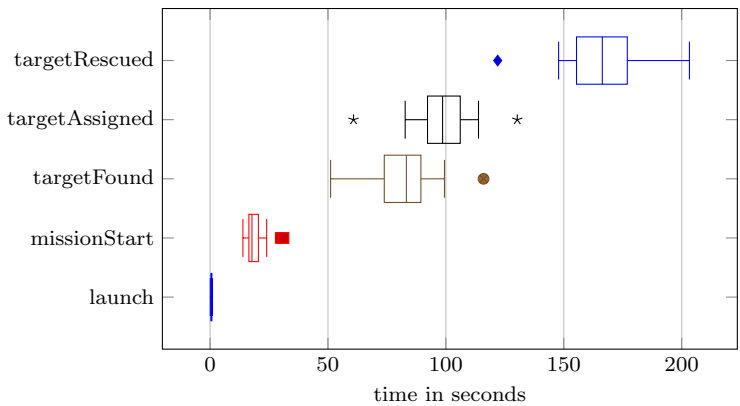

As a final step, real world experiments are performed to demonstrate that the engineering process generates code that can run on actual \acp{CPS}. For this purpose, we use an arena of size $\SI{7}{\meter} \times \SI{2.85}{\meter}$ that contains three targets at random locations. The targets are implemented as AprilTags~\cite{Wang2016} being placed on the ground. The \ac{SAR} is performed by three \acp{CPS}, one \ac{UAV} that performs the coverage and tracking and two \acp{UGV} which perform the rescue. The \acp{UGV} are placed on one side of the environment whereas the \ac{UAV} is placed at a random location inside the environment. This setup is shown in Figure~\ref{fig:arena}.
Several experiments are performed to demonstrate that the \ac{CPS} behavior runs stable and performs as expected. An exemplary execution of the experiment can be seen on YouTube\footnote{\url{https://www.youtube.com/watch?v=p6mveXV4kv0}}. Figure~\ref{fig:event_time_demo} shows the timeline of the demonstration which is averaged over 21 runs.
It shows box plots of the events happening during the mission execution. Each box is defined by the lower and upper quartiles and additionally visualizes the median, the lowest and highest datum still within 1.5 times the \ac{IQR} of the box as whiskers, and outliers and individual points. It can be seen that, despite some variations, the sequence of actions generated from the models can be successfully executed repeatedly on \acp{CPS} in real-world demonstrations. This final verification step attests the applicability of the engineering process.

\section{Conclusion}
\label{sec:conclusion}

This paper presents a toolchain for model-based engineering of \ac{CPS} swarms and provides a proof-of-concept for a \ac{SAR} scenario. We use a bottom-up design approach that yields the desired swarm behavior. We introduce a hierarchical model library that is based on the concepts of \ac{FSM} from \ac{UML} and \ac{SysML}. These models are processed by an automatic code generator that generates a software bundle ready to be compiled, deployed, and interpreted on the target \ac{CPS}.

We perform three validation and verification steps for the implementation of the \ac{CPS} swarm in a \ac{SAR} application. They serve as a proof-of-concept of the model-based engineering approach. First, we perform abstract simulations in Netlogo for rapid prototyping. Second, we perform detailed simulations in Gazebo that prove the correctness of the results. Third, we perform a physical deployment on a heterogeneous swarm of \acp{CPS}. The latter two use the generated code and run on \ac{ROS}. A limitation of the presented implementation is the high demand for computing power by the detailed simulations. This limits the applicability to simple scenarios. We envision for future improvements to overcome this issue by integrating simulation software on cloud computing clusters.

The presented tools and libraries are part of the CPSwarm workbench\footnote{\url{https://www.cpswarm.eu/index.php/cpswarm-workbench/}}. It provides a central launcher that guides the user through the engineering steps. The CPSwarm workbench furthermore provides tools for bulk deployment~\cite{Tavakolizadeh2019}, monitoring and control of the swarm, and an optimization environment that allows the generation of swarm member behaviors using evolutionary design~\cite{Sende2019}. This workbench consists of individual tools using standardized interfaces. In the future we envision adding more tools to this workbench to provide alternatives for each step of the process. This would allow to satisfy the specific requirements of different applications. Currently, there are still some steps to be performed manually. The vision is to automate more and more steps of the engineering process. We encourage the reader to contribute tools to the workbench and enhance the modeling and behavior libraries. The currently available components are published online on GitHub and as \ac{ROS} packages. They will be further developed and extended in the future. Our aim is to realize a standardized engineering approach for \ac{CPS} swarm applications.

To conclude, we can say that the CPSwarm workbench effectively improves the development process of \ac{CPS} swarms and produces efficient swarm algorithms. This paper presents the application to a \ac{SAR} scenario, however the workbench can be applied to many other scenarios such as automotive \acp{CPS}~\citep{Schranz2019a} or swarm logistics~\citep{Soriano2020}.

% *** BIBLIOGRAPHY ***
%
\bibliographystyle{spmpsci}
\bibliography{main}

\end{document}